\documentclass[twocolumn,secnumarabic,amssymb, nobibnotes, aps, prb, superscriptaddress]{revtex4-2}

\usepackage{graphicx}
\usepackage{xcolor}
\usepackage{amsmath}
\usepackage{comment}
\usepackage{afterpage}
\usepackage{placeins}
\usepackage{graphicx}
\usepackage{wrapfig}
\usepackage{soul}
\usepackage{float}
\usepackage{booktabs}
\usepackage{multirow}
\usepackage{array}
\usepackage{setspace}
\graphicspath{{Figs/}}
\usepackage{siunitx}
\usepackage{hhline}
\usepackage{xfrac}
\usepackage{graphicx}
\usepackage{dcolumn}
\usepackage{bm}
\usepackage{color,soul}
\usepackage{hyperref} 
\usepackage{here}
\usepackage{physics}
\usepackage{csquotes}

\DeclareUnicodeCharacter{0334}{~}
\DeclareUnicodeCharacter{2212}{-}

\begin{document}

\title{Anti-site disorder and Berry curvature driven anomalous Hall effect in spin gapless semiconducting Mn$_2$CoAl Heusler compound}

\author{Nisha Shahi}
\affiliation{School of Materials Science and Technology, Indian Institute of Technology (Banaras Hindu University), Varanasi 221005, India}

\author{Ajit K. Jena}
\affiliation{Indo-Korea Science and Technology Center (IKST), Bangalore 560065, India}

\author{Gaurav K. Shukla}
\affiliation{School of Materials Science and Technology, Indian Institute of Technology (Banaras Hindu University), Varanasi 221005, India}

\author{Vishal Kumar}
\affiliation{School of Materials Science and Technology, Indian Institute of Technology (Banaras Hindu University), Varanasi 221005, India}

\author{Shivani Rastogi}
\affiliation{School of Materials Science and Technology, Indian Institute of Technology (Banaras Hindu University), Varanasi 221005, India}

\author{K. K. Dubey}
\affiliation{School of Materials Science and Technology, Indian Institute of Technology (Banaras Hindu University), Varanasi 221005, India}

\author{Indu Rajput}
\affiliation{UGC-DAE Consortium for Scientific Research, Indore, India}

\author{Sonali Baral}
\affiliation{UGC-DAE Consortium for Scientific Research, Indore, India}

\author{Archana Lakhani}
\affiliation{UGC-DAE Consortium for Scientific Research, Indore, India}

\author{Seung-Cheol Lee}
\affiliation{Indo-Korea Science and Technology Center (IKST), Bangalore 560065, India}

\author{Satadeep Bhattacharjee}
\affiliation{Indo-Korea Science and Technology Center (IKST), Bangalore 560065, India}

\author{Sanjay Singh}
\affiliation{School of Materials Science and Technology, Indian Institute of Technology (Banaras Hindu University), Varanasi 221005, India}

\begin{abstract}

Spin gapless semiconductors exhibit a finite band gap for one spin channel and closed gap for other spin channel, emerged as a new state of magnetic materials with a great potential for spintronic applications. The first experimental evidence for the spin gapless semiconducting behavior was observed in an inverse Heusler compound Mn$_2$CoAl. Here, we report a detailed investigation of the crystal structure and anomalous Hall effect in the Mn$_2$CoAl using experimental and theoretical studies. The analysis of the high-resolution synchrotron x-ray diffraction data shows anti-site disorder between Mn and Al atoms within the inverse Heusler structure. The temperature-dependent resistivity shows semiconducting behavior and follows Mooij's criteria for disordered metal. Scaling behavior of the anomalous Hall resistivity suggests that the anomalous Hall effect in the Mn$_2$CoAl is primarily governed by intrinsic mechanism due to the Berry curvature in momentum space. The experimental intrinsic anomalous Hall conductivity (AHC) is found to be $\sim$35 S/cm, which is considerably larger than the theoretically predicted value for ordered Mn$_2$CoAl. Our first-principle calculations conclude that the anti-site disorder between Mn and Al atoms enhances the Berry curvature and hence the value of intrinsic AHC, which is in a very well agreement with the experiment.
 
\end{abstract}
\maketitle
\section{Introduction}

The Hall effect is defined as the generation of a transverse voltage across a current-carrying conductor, when an external magnetic field is applied perpendicular to the direction of the current flow \cite{book, nagaosa2010, smejkal2022, friedmann2013}. An additional contribution in the transverse voltage was observed in materials with broken time-reversal symmetry (TRS), known as anomalous Hall effect (AHE) that arises due to the mutual interaction of magnetization and spin-orbit coupling (SOC) \cite{nagaosa2010, golod2013}. The AHE was discovered more than a century ago, attained a vast interest in current years due to its role in understanding the fundamental physics \cite{nagaosa2010, tian2009, yu2010, nayak2016, chang2013, yang2020, taniguchi2015} and potential for applications in spintronics-based data storage devices and Hall sensors \cite{bauer2012, kim2022, Jungwirth2016, smejkal2018, bader2010, gambino1976, gerber2007, moritz2008}. Theoretical studies suggest that the AHE could originate from both extrinsic and intrinsic mechanisms \cite{tian2009, karplus1954, onoda2006}. The extrinsic mechanism receives contribution from skew scattering and side jump mechanisms that are related to the asymmetric scattering and transverse shift of propagation direction of spin-polarized charge carriers, respectively \cite{smit1958, berger1970}. On the other hand, the intrinsic mechanism is related to the momentum space Berry curvature associated with the electronic band structure \cite{karplus1954,luttinger1958, berry1984, wang2007}. The momentum space Berry curvature acts as a fictitious magnetic field and introduces an anomalous velocity in addition to the group velocity of the electronic wave, perpendicular to the electric field direction, which creates intrinsic anomalous Hall conductivity (AHC) in the system \cite{nagaosa2010, karplus1954, berry1984}.

The large AHC due to non-vanishing momentum space Berry curvature has been observed in various kinds of systems such as  Nd$_2$Mo$_2$O$_7$ \cite{taguchi2001}, (In, Mn)As \cite{jungwirth2002}, (Ga, Mn)As \cite{jungwirth2002}, Mn$_3$Sn \cite{nakatsuji2015}, Mn$_3$Ge \cite{nayak2016}, Fe$_3$Sn$_2$ \cite{wang2016}, Co$_3$Sn$_2$S$_2$ \cite{wang2018}, CoNb$_3$S$_6$ \cite{ghimire2018}, and SrIrO$_3$ \cite{yoo2021}. Besides these materials, Heusler alloys emerged as promising candidate for the realization of large Berry curvature originated from their peculiar band structure due to integrated effect of crystal symmetry and SOC \cite{ernst2019, manna2018, Li2020, gaurav2021, Noky2020}. It has been observed that any small perturbation in the electronic band structure for example a local disorder may affect the momentum space Berry curvature and therefore the intrinsic AHC in the system \cite{vidal2011, hazra2018, wang2019, kastbjerg2013, shen2020}. Recently, an enhanced intrinsic AHC has been observed in Fe$_2$-based inverse Heusler compounds due to the anti-site disorder \cite{mende2021}. Also, a Berry curvature driven enhanced intrinsic AHC has been observed in Co$_2$FeAl Heusler compound, originated via anti-site disorder between Fe and Al atoms \cite{gauravc2021}. 

The discovery of the spin gapless semiconducting (SGS) behavior in the Mn$_2$CoAl Heusler compound puts this material forward as an important candidate for  technological application in the field of spintronics \cite{ouardi2013, wang2008, kim2022, bader2010}. Experimental studies on thin film as well as bulk systems of Mn$_2$CoAl reflect that most of the systems crystallize with compositional and/or anti-site disorder and the reported value of AHC is not in an agreement with the theory \cite{ouardi2013, arima2018, buckley2019, xu2019, xu2014, jamer2013, chen2018, feng2015, sun2016, Xin2017, zhang2013, ueda2017}. A theoretical investigation considering anti-site disorder suggests an enhancement in AHC, necessitates a detailed  experimental and theoretical investigation to understand the origin of AHC in Mn$_2$CoAl Heusler compound \cite{kudrnovsky2013}. 

In the present manuscript, we investigate AHE in the Mn$_2$CoAl Heusler compound experimentally as well as theoretically. The high-resolution synchrotron x-ray diffraction (SXRD) analysis exhibits 25\% anti-site disorder between Mn and Al atoms. The scaling behavior of anomalous Hall resistivity reveals the major contribution of intrinsic mechanism in the observed AHE. Experimentally, we found a higher value of intrinsic AHC as compared to the value reported for an ordered Mn$_2$CoAl from theory \cite{kudrnovsky2013, ouardi2013}. Our theoretical calculations show that the anti-site disorder between Mn and Al atoms enhances the Berry curvature and hence the value of intrinsic AHC, which is in well agreement with the experiment.

\section{Methods}

\begin{figure*}[t]
\centering  
\includegraphics[width=1\linewidth]{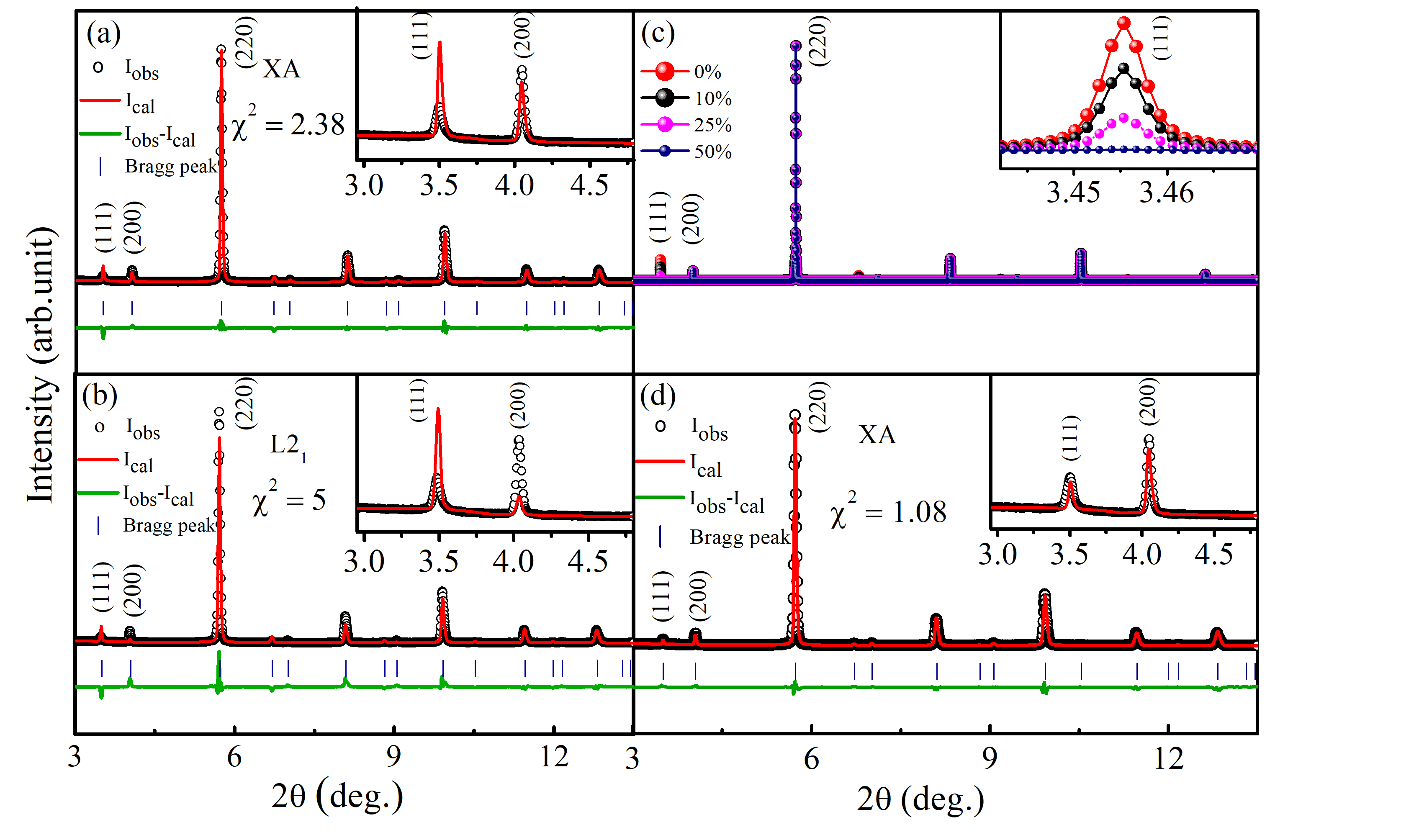}
\caption{(a) Rietveld profile fitting of the room temperature (RT) synchrotron x-ray diffraction (SXRD) pattern of  Mn$_2$CoAl compound by considering the XA-type ordered structure. The observed (I\textsubscript{obs}), calculated (I\textsubscript{cal}) and difference between observed and calculated profiles (I\textsubscript{obs}-I\textsubscript{cal}) are shown by black circle, red continuous line and green continuous line, respectively. The blue tick bars and the $\chi^2$ value indicate the Bragg peak positions and the goodness of fit, respectively. (b) Rietveld profile fitting of the RT SXRD pattern of  Mn$_2$CoAl compound by considering the L2$_1$ type ordered structure. Insets in (a) and (b) show an enlarged view of (111) and (200) peaks. (c) Simulated SXRD patterns considering the different percentage of anti-site disorder between Mn\textsubscript{4c} and Al atoms. Inset shows the variation in (111) peak intensity with the different percentage of anti-site disorder. (d) Rietveld profile fitting of the RT SXRD pattern of  Mn$_2$CoAl by considering 25\% anti-site disorder between  Mn\textsubscript{4c} and Al atoms. Inset shows an enlarged view around (111) and (200) peaks.}
\label{fig:XRD}
\end{figure*}

Polycrystalline Mn$_2$CoAl Heusler compound is prepared by the arc-melting method \cite{holt1978, sanjay2017, sanjay2012}. The constituent elements (purity higher than 99.9\%) of the intermetallic system are melted in a water-cooled copper hearth under argon atmosphere (purity better than 99.999\%). The sample is remelted 5 times to ensure the homogeneous mixing of the constituents. The chemical composition is verified using the energy dispersive analysis of x-ray technique. The average composition is found to be Mn$_{1.99}$Co$_{1.00}$Al$_{0.99}$, which corresponds to Mn$_2$CoAl. A small piece of the sample is powdered and room temperature SXRD is performed in PETRA III DESY, Germany (wavelength $\sim$0.207 \si{\angstrom}) for structural analysis. The direct current (DC) magnetization measurements are performed using a 9 Tesla (T) physical property measurement system (PPMS) of quantum design(QD). Transport measurements are carried out using alternating current transport (ACT) option of 9 T PPMS of QD. Resistivity and magneto-resistance measurements are performed by the four-probe method, whereas the five-probe method is used for the Hall measurement. To remove the longitudinal resistivity contribution in the Hall data due to voltage probe misalignment, we have anti-symmetrized the Hall resistivity data by using the formula \si{\rho}\textsubscript{\tiny{H}} = [\si{\rho}\textsubscript{\tiny{H}}(+H) - \si{\rho}\textsubscript{\tiny{H}}(-H)]/2.

The spin-polarized Kohn-Sham Hamiltonian was solved within the framework of pseudo-potentials (PP) and plane waves as implemented in Quantum ESPRESSO (QE) density-functional theory (DFT) package \cite{giannozzi2009}. Exchange-correlation part of the above Hamiltonian is approximated by generalized gradient approach \cite{perdew1996} through ONCV Vanderbilt PPs \cite{hamann2013}. The kinetic energy cutoff of 80 Ry is used to fix the number of plane waves. The same Gaussian smearing value (0.01 Ry) is used both for the self-consistent (SC) and non-self-consistent (NSC) calculations to carry out electronic integration over the Brillouin zone (BZ). A tight energy threshold ($10^{-8}$ Ry) is considered for the SC energy calculations. We have used the WANNIER90 tool (implemented within QE) in order to compute the Wannier interpolated bands, Berry curavture and AHC \cite{giannozzi2009, marzari2001, souza2001, pizzi2020}. We note that spin-orbit coupling (SOC) is introduced in all the AHC related calculations. Collinear spin-polarized calculations are also performed to compare some of our results with earlier reports and are mentioned at the appropriate places. The Monkhorst-Pack \textbf{k}-grid of $8\times8\times8$ of the BZ is considered in all the calculations, viz.: SC, NSC, and WANNIER90. We found that the use of transition metal-$d$ orbitals as the projections in the WANNIER90 calculations provide the good interpolation. A denser BZ \textbf{k}-grid of $75\times75\times75$ is taken to calculate the intrinsic AHC. Through the  adaptive refinement technique, a further fine mesh of $5\times5\times5$ is added around the points wherever the mode of the Berry curvature ($\abs{\Omega(\textbf{k})}$) exceeds 100 bohr$^2$. The calculations are carried out using experimental refined lattice constant of 5.858 \AA.

\section{Results and discussion}

\def\thesubsection{\alph{subsection}}
\subsection{Structural characterization}

To investigate the crystal structure of the Mn$_2$CoAl Heusler compound, the SXRD pattern was collected at room temperature. Mn$_2$YZ  Heusler compounds generally exhibit XA-type (prototype Hg$_2$CuTi, space group \textit{F{\si{\overline{4}}}}3m) crystal structure, if atomic number of Y is higher than the Mn like Mn$_2$CoAl \cite{graf2011}. Therefore, in the primary step, we performed Rietveld refinement of SXRD data by assuming the XA-type crystal structure (Fig.\,\ref{fig:XRD}(a)), using FULLPROF software package \cite{full}. In the refinement, the  Mn atoms were considered at 4a (0, 0, 0) and 4c (0.25, 0.25, 0.25) Wyckoff positions and will be denoted as Mn\textsubscript{4a} and Mn\textsubscript{4c}, respectively. The Co and Al atoms were considered at 4b (0.50, 0.50, 0.50) and 4d (0.75, 0.75, 0.75) Wyckoff positions, respectively. It can be noticed from Fig.\,\ref{fig:XRD}(a) that the observed and calculated peak profiles do not match with each other as the calculated intensity of (111) superlattice peak is much larger than the observed intensity (depicted in the inset of Fig.\,\ref{fig:XRD}(a)). This mismatch between the calculated and observed intensities of Bragg peaks indicates that Mn$_2$CoAl does not crystallize in an ordered XA-type structure. In the next step, we tried the Rietveld refinement with the ordered L2$_1$ type (\textit{Fm{\si{\overline{3}}}m}) crystal structure (Fig.\,\ref{fig:XRD}(b)) \cite{sanjay2012, graf2011}, which also fails to account the intensities of both the (111) and (200) superlattice reflections as depicted in the Fig.\,\ref{fig:XRD}(b) and its inset.

 In literature it has been suggested that the most of the inverse Heusler compounds crystallize with anti-site disorder \cite{nayak2015, nayak2013, graf2011, mende2021, sanjay2012}. The anti-site disorder in the Heusler compounds can be generally captured by analyzing the intensities of ordering dependent (111) and (200) superlattice reflections \cite{dliu2008}. For Mn$_2$CoAl compound, the structure factor for (111), (200) and (220) reflections can be written as- 
 \begin{equation}
 F_{111} = 4[(f\textsubscript{Mn\textsubscript{4a}} - f\textsubscript{Co}) - i(f\textsubscript{Mn\textsubscript{4c}} - f\textsubscript{Al})]   
 \end{equation} 
 
  \begin{equation}
 F_{200} = 4[(f\textsubscript{Mn\textsubscript{4a}} + f\textsubscript{Co}) - (f\textsubscript{Mn\textsubscript{4c}} + f\textsubscript{Al})]   
 \end{equation} 
 
  \begin{equation}
 F_{220} = 4[(f\textsubscript{Mn\textsubscript{4a}} + f\textsubscript{Co}) + (f\textsubscript{Mn\textsubscript{4c}} + f\textsubscript{Al})]   
 \end{equation} 
 
 Eq.\,(1) suggests that the intensity of (111) superlattice reflection is due to the difference between the atomic scattering factor (SF) of Mn\textsubscript{4c} and Al atoms as the difference between the atomic SF of Mn\textsubscript{4a} and Co atoms is negligible (since both are neighbouring elements in the periodic table) and also the intensity of (111) peak may change, if the system has the anti-site disorder between the Mn\textsubscript{4c} and Al atoms. From Eq.\,(2) and Eq.\,(3), it is obvious that the intensities of (200) and (220) reflection will be unaffected from the anti-site disorder between the Mn\textsubscript{4c} and Al atoms. Hence, the mismatch between the observed and calculated intensities of (111) peak in the Rietveld fitted SXRD pattern with XA-type crystal structure suggests the presence of anti-site disorder between Mn\textsubscript{4c} and Al atoms.

Now, to find out the amount of anti-site disorder, we simulated the SXRD pattern using VESTA software \cite{momma2011} by incorporating the different percentage of anti-site disorder between Mn\textsubscript{4c} and Al atoms within XA-type crystal structure as shown in Fig.\,\ref{fig:XRD}(c). The inset of Fig.\,\ref{fig:XRD}(c) represents an enlarged view of the change in intensity of (111) superlattice reflection with anti-site disorder. We found that for 25\% anti-site disorder between Mn\textsubscript{4c} and Al atoms, the intensity ratio of (111) superlattice reflection to the (220) fundamental reflection is in good agreement with the experiment. It is worthwhile to mention here that SXRD patterns simulated using considering other types of disorder like disorder between Mn\textsubscript{4a} and Mn\textsubscript{4c}, between Mn\textsubscript{4a} and Al and between Co and Al do not provide intensity ratio of Bragg reflections as observed in experimental SXRD pattern. Therefore, based on the information obtained from the Eq.\,(1-3) and simulated SXRD patterns, in the final step we performed the Rietveld refinement considering the 25\% anti-site disorder between the Mn\textsubscript{4c} and Al atoms. A very good match between the observed and calculated peak profile was observed  as depicted in Fig\,\ref{fig:XRD}(d). The inset of Fig.\,\ref{fig:XRD}(d) shows an enlarged view of (111) and (200) peaks. The refined lattice parameter was obtained $\sim$5.858 \si{\angstrom}, which is in well agreement with the literature \cite{xu2019}.
 
 \def\thesubsection{\alph{subsection}}
\subsection{Magnetization and resistivity measurements}

\begin{figure*}[t]
\centering  
\includegraphics[width=1\linewidth]{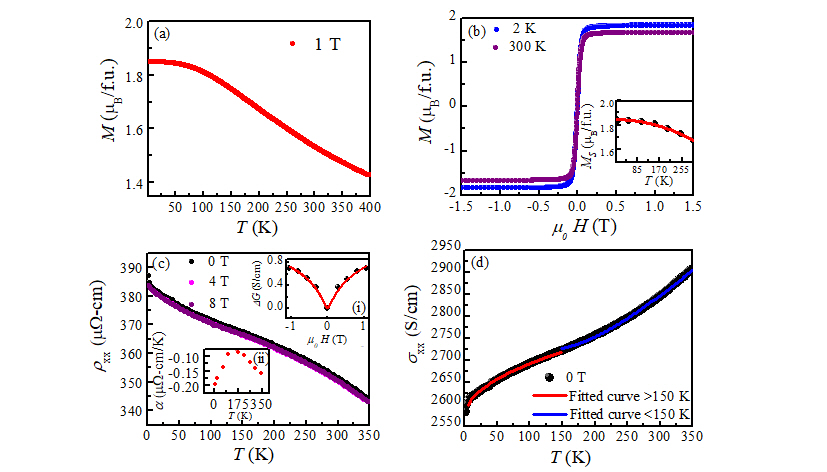}
\caption{(a) Temperature-dependent magnetization curve at 1 T magnetic field. (b) The isothermal field-dependent magnetization at different temperatures. Inset shows the fitting of  saturation magnetization (\textit{M\textsubscript{s}}) vs temperature (\textit{T}) with empirical relation M\textsubscript{s}=M\textsubscript{0}[1-(T/Tc)$^2$]$^{1/2}$. The black dots and the red continuous line in the inset represent the \textit{M\textsubscript{s}} vs \textit{T} data and fitted curve, respectively. (c) Resistivity (\si{\rho}\textsubscript{xx}) vs temperature (\textit{T}) plot at different magnetic fields. Insets (i) and (ii) show the fitting of magneto-conductivity with field by Hikami-Larkin-Nagaoka equation and the variation of temperature coefficient of resistivity (\si{\alpha}) with temperature (\textit{T}), respectively. (d) Fitting of conductivity (\si{\sigma}\textsubscript{xx}) vs temperature (\textit{T}) data in the temperature range below 150 K and in higher temperature range above 150 K at zero field.}
\label{fig:MT}
\end{figure*}

The temperature variation of magnetization measured in a temperature range of 2-400\,K under the magnetic field of 1\,T, shown in Fig.\,\ref{fig:MT}(a). The magnetization decreases with increasing temperature, as expected for ferrimagnetic system \cite{bezmat2014}. The observed behavior is similar to the earlier reports \cite{xu2019, ouardi2013}. Fig.\,\ref{fig:MT}(b) presents the field-dependent magnetic isotherms (\textit{M(H)} curves) at 2\,K and 300\,K. The negligible hysteresis is due to the soft magnetic nature of the Mn$_2$CoAl compound. The \textit{M(H)} curves show similar behavior in the entire temperature range of 2-300\,K. To calculate the Curie temperature (T$_c$) of present compound, the saturation magnetization (\textit{M\textsubscript{s}}) vs temperature (\textit{M$_{s}$(T)}) data obtained from the \textit{M(H)} curves (black spheres in the inset of Fig.\,\ref{fig:MT}(b)) were fitted by the  empirical law M\textsubscript{s}= M\textsubscript{0}[1-(T/Tc)$^2$]$^{1/2}$ as shown by the red curve in the inset of Fig.\,\ref{fig:MT}(b) \cite{xu2014}. From this fitting, the T$_c$ was found to be 720\,K, which is in well agreement with the value reported in the literature \cite{xu2014, xu2019, ouardi2013}. The saturation magnetization of present compound is found to be \si{\sim}1.9 \si{\mu}B/f.u. at 2\,K, which 
is in well agreement with the literature \cite{ouardi2013,xu2014}.

The monotonous decrease in the longitudinal resistivity (\si{\rho}\textsubscript{xx}) with temperature (Fig.\,\ref{fig:MT}(c)) illustrates the semiconducting behavior of the present compound. A slight decrease in the value of \si{\rho}\textsubscript{xx} with increasing magnetic field indicates the presence of small negative magnetoresistance in the system. The resistivity at 2\,K (that corresponds to residual resistivity) is \si{\sim}387 \si{\mu}\si{\ohm}-cm, which is comparable to the reported one \cite{xu2019}. The residual resistivity ratio [\si{\rho}\textsubscript{xx}(300 K)/\si{\rho}\textsubscript{xx}(2 K)] was found \si{\sim} 0.92, which is comparable to the value in the literature \cite{xu2019, ouardi2013} and also suggest the presence of disorder in the Mn$_2$CoAl compound. 

A theoretical calculation  considering anti-site disorder into account predicts a half-metallic character of Mn$_2$CoAl Heusler compound in contrast to the SGS behavior \cite{kudrnovsky2013, galanakis2014}. It is interesting to note that the Mn$_2$CoAl compound shows the semiconducting behavior (Fig.\,\ref{fig:MT}(c)) although a large (25\%) anti-site disorder is observed. It has been suggested in literature that if the system has weak localization (WL) then inelastic scattering increases with increasing the temperature, which may destroy the phase coherence, and hence the resistivity decreases with increasing temperature \cite{dugdale1995}. To check the possibility of WL in Mn$_2$CoAl, the magneto-conductivity (MC) i.e. difference of conductivity between finite magnetic field  and zero field at room temperature is fitted by the Hikami-Larkin-Nagaoka (HLN) equation \cite{hikami1980}. Although the HLN model explains the WL effect in two-dimensional systems but it has been observed that the several three dimensional (3D) systems also follow this model \cite{xu2014, Singh2022, chamorro2019, sasmal2020, laha2019, chen2020}. The HLN equation is given as-

\begin{equation}
\si{\Delta}G(H) = \frac{\si{\gamma}e^2}{\si{\pi}h}\left[\si{\psi}\left(\frac{h}{4eHL\textsubscript{\si{\phi}}^2}+\frac{1}{2}\right) - ln\left(\frac{h}{4eHL\textsubscript{\si{\phi}}^2}\right)\right]
\end{equation}

where \textit{\si{\Delta}G} is MC, \si{\psi} is the digamma function, \textit{L}\textsubscript{\si{\phi}} is the phase coherence length, \textit{e} is the electronic charge, \textit{h} is the Planck’s constant, \textit{H} is the magnetic field, and \si{\gamma} is a prefactor. The sign of prefactor \si{\gamma} decides whether there is WL or weak anti-localization in the system. The positive MC (black dots in the inset (i) of Fig.\,\ref{fig:MT}(c)), fitted (red solid line Fig.\,\ref{fig:MT}(c)) by above equation with positive \si{\gamma} and a large logarithmic increment for lower magnetic field region indicates the presence of WL in Mn$_2$CoAl \cite{dulal2019}. Moreover, we may see that the magnetoconductivity in the low field region follows H$^{0.5}$ power law variation, which is well established for WL in 3D systems \cite{kawabata1980, baxter1989, chen2019, malick2022, afzal2020}, also validate the presence of WL in bulk Mn$_2$CoAl. Further, Mooij established a correlation between temperature coefficient of resistivity and residual resistivity for disordered metal which describes that a disordered metal may exhibit semiconducting behavior, if the residual resistivity exceeded from 150 \si{\mu}\si{\ohm}-cm \cite{mooij1973}. For the Mn$_2$CoAl, a larger value of the  residual resistivity (\si{\sim}387 \si{\mu}\si{\ohm}-cm) is observed and follows the Mooij's criteria.

The temperature coefficient of resistivity (\si{\alpha}) for Mn$_2$CoAl changes from -0.9\si{\times}10$^{-9}$ to -1.9\si{\times}10$^{-9}$ \si{\si{\ohm}}-m/K in the temperature range of 2-350 K  with a slope change around 150\,K (shown in the inset (ii) of Fig.\,\ref{fig:MT}(c)). To understand the origin of slope change in \si{\alpha}, we analyzed the longitudinal conductivity (\si{\sigma}\textsubscript{xx} = 1/\si{\rho}\textsubscript{xx}) vs temperature data in the absence of magnetic field, and we found that \si{\sigma}\textsubscript{xx} is proportional to the T$^{1/2}$ below the temperature 150 K as shown in Fig.\,\ref{fig:MT}(d). This conductivity behavior follows the interaction theory at low-temperature according to Kaveh and Mott \cite{kaveh1982}, which might be due to  anti-site disorder as supported by our SXRD analysis. Above the 150 K, the conductivity variation with temperature follows the semiconducting-like character and is well approximated by the relation (\si{\sigma}\textsubscript{xx} = \si{\sigma}\textsubscript{0} + \si{\sigma}\textsubscript{a} exp$^{−E_g/K_BT}$) as shown in Fig.\,\ref{fig:MT}(d) \cite{gofryk2005}. The calculated band gap (E$_g$) from the best fitting was found to be \si{\sim}50 meV, that is of the same order as reported for the thin film of Mn$_2$CoAl \cite{xu2014}. Recently, the appreciable variation in Seebeck effect around 150 K has been observed in the polycrystal Mn$_2$CoAl \cite{ouardi2013}. This shows that our prepared compound is similar to the reported one and the disorder present in Mn$_2$CoAl compound is inherent and does not change with the sample preparation conditions.


\def\thesubsection{\alph{subsection}}
\subsection{Anomalous Hall measurement}

\begin{figure*}[t]
\centering  
\includegraphics[width=1\linewidth]{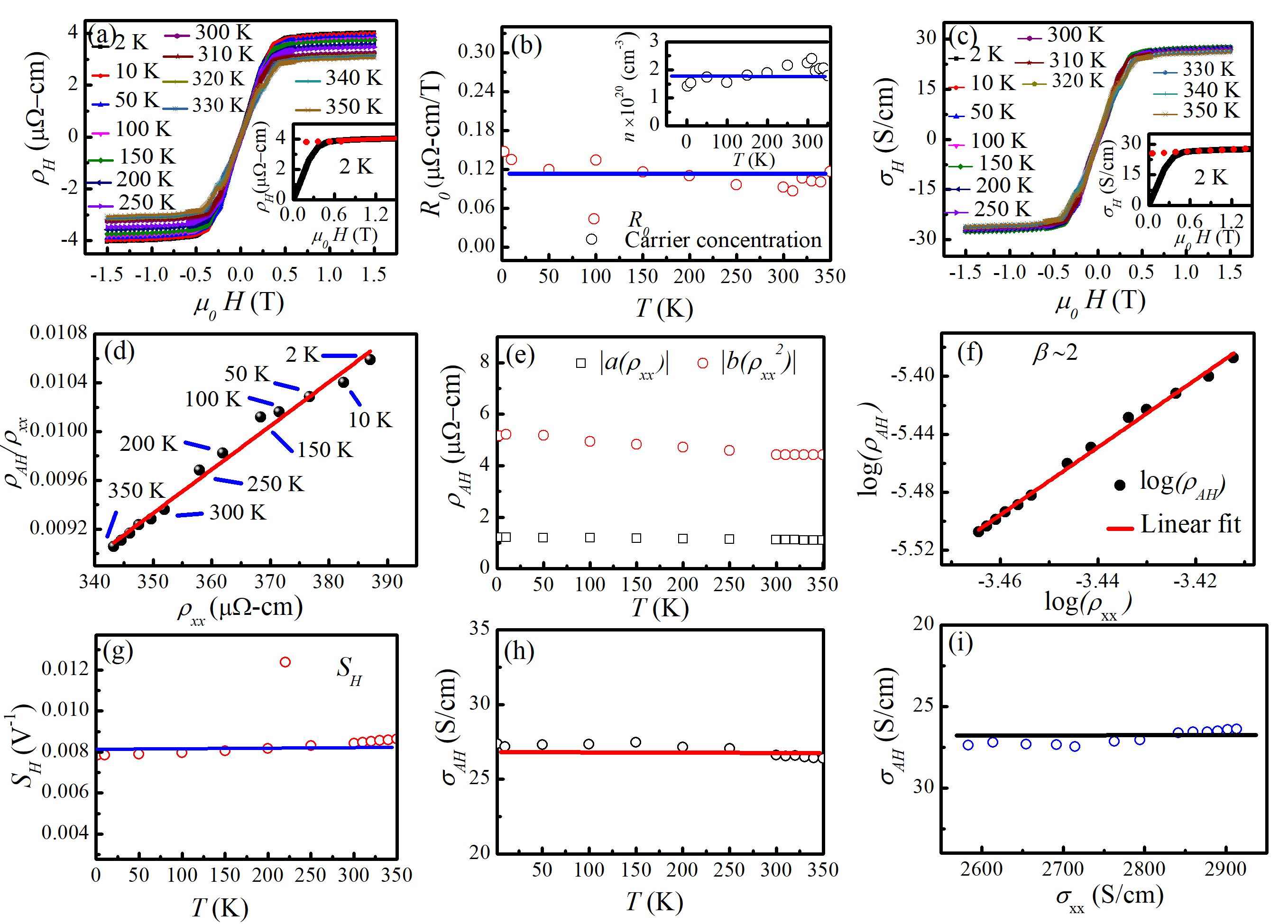}
\caption{(a) Field-dependent Hall resistivity curves at different temperatures and the inset shows the fitting of Hall data in higher field region $>$ 0.6 T at 2 K. (b) Variation of ordinary Hall coefficient with temperature and the inset shows carrier concentration vs temperature plot. (c) Field-dependent Hall conductivity curves at different temperatures. The inset indicates zero field extrapolation of high field Hall conductivity data at 2 K, shown by red dotted line. (d) Fitting of ratio of anomalous Hall resistivity and longitudinal resistivity (\si{\rho}\textsubscript{\tiny{AH}}/\si{\rho}\textsubscript{xx}) vs \si{\rho}\textsubscript{xx} data. (e) Different contributions in \si{\rho}\textsubscript{\tiny{AH}} with temperature are plotted on the same scale. (f) Linear fitting of log(\si{\rho}\textsubscript{\tiny{AH}}) vs log(\si{\rho}\textsubscript{xx}) data. (g) Variation of anomalous Hall scaling coefficient (\textit{S\textsubscript{H}}) with temperature. (h) Variation of AHC with temperature. (i) Variation of AHC with longitudinal conductivity.}
\label{fig:AHR}
\end{figure*}

Hall measurements were carried out in the temperature range of 2-350\,K to investigate the anomalous transport behavior of Mn$_2$CoAl compound. In general the total Hall resistivity (\si{\rho}\textsubscript{H}) consists of two parts namely ordinary Hall and anomalous Hall and can be written as \cite{book, kida2011}-
\begin{equation}
\si{\rho}\textsubscript{H} = {R_0}H + {R_s}{M_s}
\end{equation}

where \textit{R$_0$} and \textit{R$_s$} are the ordinary and anomalous Hall coefficients, respectively. \textit{M$_s$} corresponds to the saturation magnetization and  \textit{R$_s$M$_s$} represents the magnitude of anomalous Hall resistivity (\si{\rho}\textsubscript{AH}).The field-dependent \si{\rho}\textsubscript{H} data were recorded at different temperatures upto magnetic field of 1.5\,T as shown in the Fig.\,\ref{fig:AHR}(a). The \si{\rho}\textsubscript{H} increases steeply upto \si{\sim} 0.6\,T field, which is observed due to AHE. At the higher field region ($>$0.6 T), \si{\rho}\textsubscript{H} changes linearly and shows positive slope with magnetic field, which is due to the ordinary Hall effect \cite{gaurav2021}. To separate out the ordinary and anomalous Hall contributions, we performed fitting of the \si{\rho}\textsubscript{H} vs \textit{H} data by using the Eq.\,(5) in the higher field region ($>$0.6 T). The fitting of the \si{\rho}\textsubscript{H} vs \textit{H} data at 2 K, shown by the continuous red line in the inset of Fig.\,\ref{fig:AHR}(a), provides the value of \textit{R$_0$} and \textit{R$_s$M$_s$} that correspond to the slope and intercept on the y axis of the fitted line. A similar analysis (i.e. fitting) is performed  at various temperatures in the temperature range of 10-350 K and Fig.\,\ref{fig:AHR} (b) shows a temperature variation of \textit{R$_0$} obtained from the fitting. The positive value of \textit{R$_0$} in whole temperature range (2-350 \,K) reveals that holes are the dominating charge carriers in the transport. The carrier concentration (\textit{n}) determined by the expression \textit{n} = $\frac{1}{eR_0}$, was found to be \si{\sim}1.5\si{\times}10$^{20}$cm$^{-3}$ and \si{\sim}2\si{\times}10$^{20}$cm$^{-3}$ at 2\,K and 350\,K, respectively. Thus, the value of \textit{n} is nearly temperature independent. The temperature variation of \textit{n} for temperature range 2-350 \,K is depicted in the inset of Fig.\,\ref{fig:AHR}(b), which is similar to as reported in the literature \cite{ouardi2013}. The Hall conductivity (\si{\sigma}\textsubscript{\tiny{H}}) has been extracted by using the equation \cite{hazra2018, manna2018}-
\begin{equation}
\si{\sigma}\textsubscript{\tiny{H}}= \frac{\si{\rho}\textsubscript{\tiny{H}}}{\si{\rho}\textsubscript{\tiny{H}}^2+\si{\rho}\textsubscript{xx}^2}
\end{equation}

The field-dependent Hall conductivity at different temperatures is shown in Fig.\,\ref{fig:AHR}(c). The value of AHC at 2\,K is determined by zero field extrapolation of the higher field Hall conductivity curve as elucidated by red dotted line in the inset of the Fig.\,\ref{fig:AHR}(c). The obtained value of AHC is 27 S/cm, which is comparable to the experimentally found value in the literature \cite{ouardi2013, xu2019}. From the Berry curvature calculations, the AHC value was reported about 3 S/cm \cite{ouardi2013}, which is an order of magnitude lower than the value obtained from the experiment (\si{\sim} 27 S/cm). To understand this discrepancy, it is necessary to address whether there is contribution in the AHC from an extrinsic mechanisms, such as skew scattering  and side jumps, or from intrinsic mechanism due to the momentum space Berry curvature associated with electronic band structure. 

To calculate the separate contributions of extrinsic and intrinsic mechanism in the total AHC, the  \si{\rho}\textsubscript{\tiny{AH}}/\si{\rho}\textsubscript{xx} vs \si{\rho}\textsubscript{\tiny{xx}} data (black dots in Fig.\,\ref{fig:AHR}(d)) was fitted (red line in Fig.\,\ref{fig:AHR}(d)) using the following relation \cite{he2012, kotzler2005, tian2009}-
\begin{equation}
 \si{\rho}\textsubscript{\tiny{AH}}/\si{\rho}\textsubscript{xx} = a + b\si{\rho}\textsubscript{xx}
 \end{equation}
where the parameters \textit{a} and \textit{b} contain information about extrinsic skew scattering and the combined effect of extrinsic side jump and intrinsic contribution, respectively. From the \si{\rho}\textsubscript{\tiny{AH}}/\si{\rho}\textsubscript{xx} vs \si{\rho}\textsubscript{\tiny{xx}} data fitting with above equation, we obtained the value of \textit{a} \si{\sim} -0.003 and \textit{b} \si{\sim} 35 ($\si{\ohm}.cm)^{-1}$. The negative value of coefficient \textit{a} indicates that extrinsic skew scattering contribution is opposite to both side jump and intrinsic contribution due to momentum space Berry curvature. The value of parameter \textit{b} contains the contributions in  AHC due to both side jump and momentum space Berry curvature. By using the coefficients \textit{a} and \textit{b}, we calculated skew scattering term (a\si{\rho}\textsubscript{xx}) and intrinsic plus side jump term ($b\si{\rho}\textsubscript{xx}{^2}$) and plotted  on the same scale as shown in Fig.\,\ref{fig:AHR}(e). We can clearly see that side jump together with intrinsic contribution dominates over the skew scattering contribution in the overall AHC in the temperature range 2-350 K.

The dominating mechanism in the AHE can be alternatively evaluated by the exponent \si{\beta} using the scaling relation \si{\rho}\textsubscript{\tiny{AH}}\si{\propto}\si{\rho}\textsubscript{xx}\textsuperscript{\si{\beta}} \cite{nagaosa2010,gauravc2021, roy2020}. If \si{\beta} = 1, the skew scattering mechanism  will dominant in AHE and if \si{\beta} = 2, the combination of side jump and intrinsic mechanism will largely contribute to the AHE. The exponent \si{\beta} determined by linear fitting of log(\si{\rho}\textsubscript{\tiny{AH}}) vs log(\si{\rho}\textsubscript{xx}) is turned out to be 2, which also supports that the AHC is dominated by the side jump and intrinsic mechanism \cite{gauravc2021, roy2020}. It is not possible practically to separate out the side jump and intrinsic contributions because both the contributions show quadratic dependency on \si{\rho}\textsubscript{xx} \cite{nagaosa2010}. However, the AHC due to side jump mechanism can be approximated using an expression (e$^2$/(ha)(E\textsubscript{so}/E\textsubscript{F}), where E\textsubscript{so} is the spin–orbit interaction energy and E\textsubscript{F} is Fermi energy \cite{nozieres1973, onoda2006}. The physical quantities e, h and a are the electronic charge, Planck's constant and lattice parameter, respectively. Here, for Mn$_2$CoAl the \si{\sigma}\textsubscript{xx} is \si{\sim}2.6\si{\cross}10$^3$ S/cm. Therefore, we may conclude that Mn$_2$CoAl is in the moderately dirty regime for which E\textsubscript{so} \si{\sim}$\hbar$/\si{\tau}, where $\hbar$ is reduced Planck's constant i.e. h/2\si{\pi} and \si{\tau} is the scattering time \cite{nagaosa2010,kim2018}. The calculated value of the ratio E\textsubscript{so}/E\textsubscript{F} for Mn$_2$CoAl is of the order of $10^{-3}$, and hence in AHC, side jump contribution is negligible as compared to the  Berry curvature induced intrinsic contribution. 
It has been suggested that the intrinsic AHC is nearly proportional to the magnetization, thus the scaling coefficient \textit{S\textsubscript{H}} = $\frac{\si{\sigma}_\textsubscript{\tiny{AH}}}{M_s}$ should show temperature independent behavior, if there is dominant intrinsic contribution to AHE \cite{wang2016, roy2020, zeng2006}. Fig.\,\ref{fig:AHR}(g) depicts that the \textit{S\textsubscript{H}} for Mn$_2$CoAl is indeed independent of temperature, which indicates that intrinsic contribution due to momentum space Berry curvature dominates over the extrinsic scattering contributions in the AHE \cite{wang2016, roy2020, zeng2006}. The AHC shows a temperature independent behavior (Fig.\,\ref{fig:AHR}(h))  and does not vary with \si{\sigma}\textsubscript{xx} as well (Fig.\,\ref{fig:AHR}(i)).  This robust behavior of the AHC with the temperature and \si{\sigma}\textsubscript{xx} further confirms that AHE in Mn$_2$CoAl is mainly  originated from the intrinsic mechanism, and is thus dominated by Berry curvature in the momentum space \cite{eliu2018}.  

\def\thesubsection{\alph{subsection}}
\subsection{First-principle calculations}

\begin{figure}[t]
\centering
\includegraphics[width=1\linewidth]{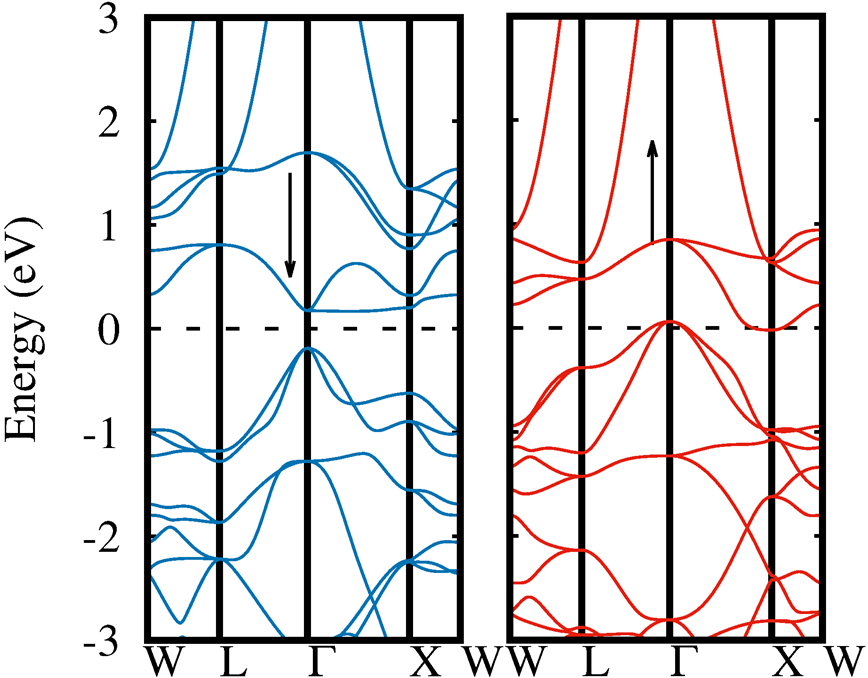}
\caption{Spin-polarized band structure of ordered Mn$_2$CoAl. Down and up arrows represent the spin-minority and majority electrons, respectively.}
\label{fig:mca-sp-wan_soc}
\end{figure}

\begin{figure}
\centering
\includegraphics[width=1\linewidth]{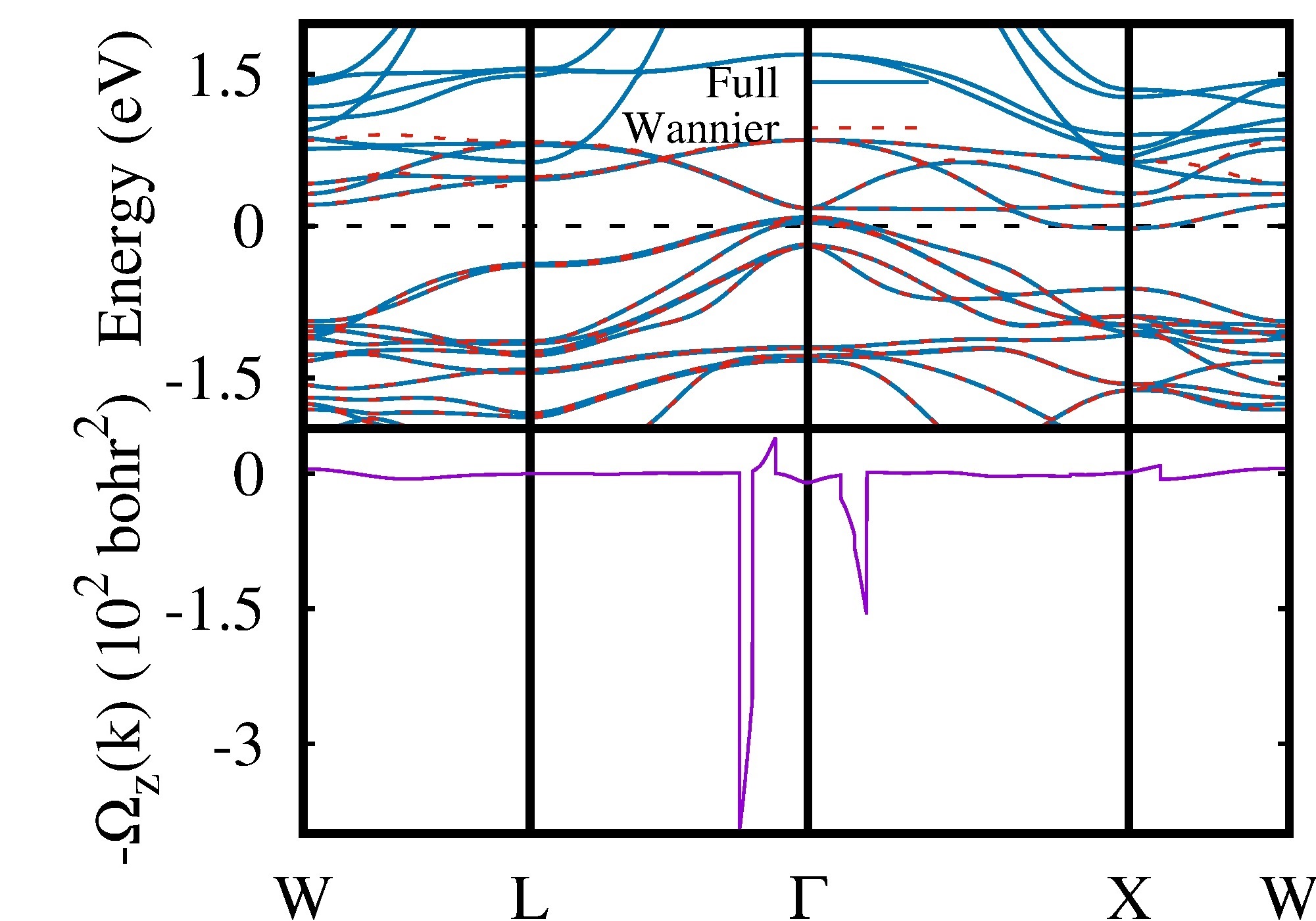}
\caption{Top: Full electronic band structure (blue continuous line) and Wannier interpolated band structure (red dashed line) of ordered Mn$_2$CoAl. The Fermi energy is set to 0 eV. Bottom: Distribution of the Berry curvature along high symmetry path in the Brillouin zone.}
\label{fig:ord}
\end{figure}

\begin{figure}[t]
\centering
\includegraphics[width=1\linewidth]{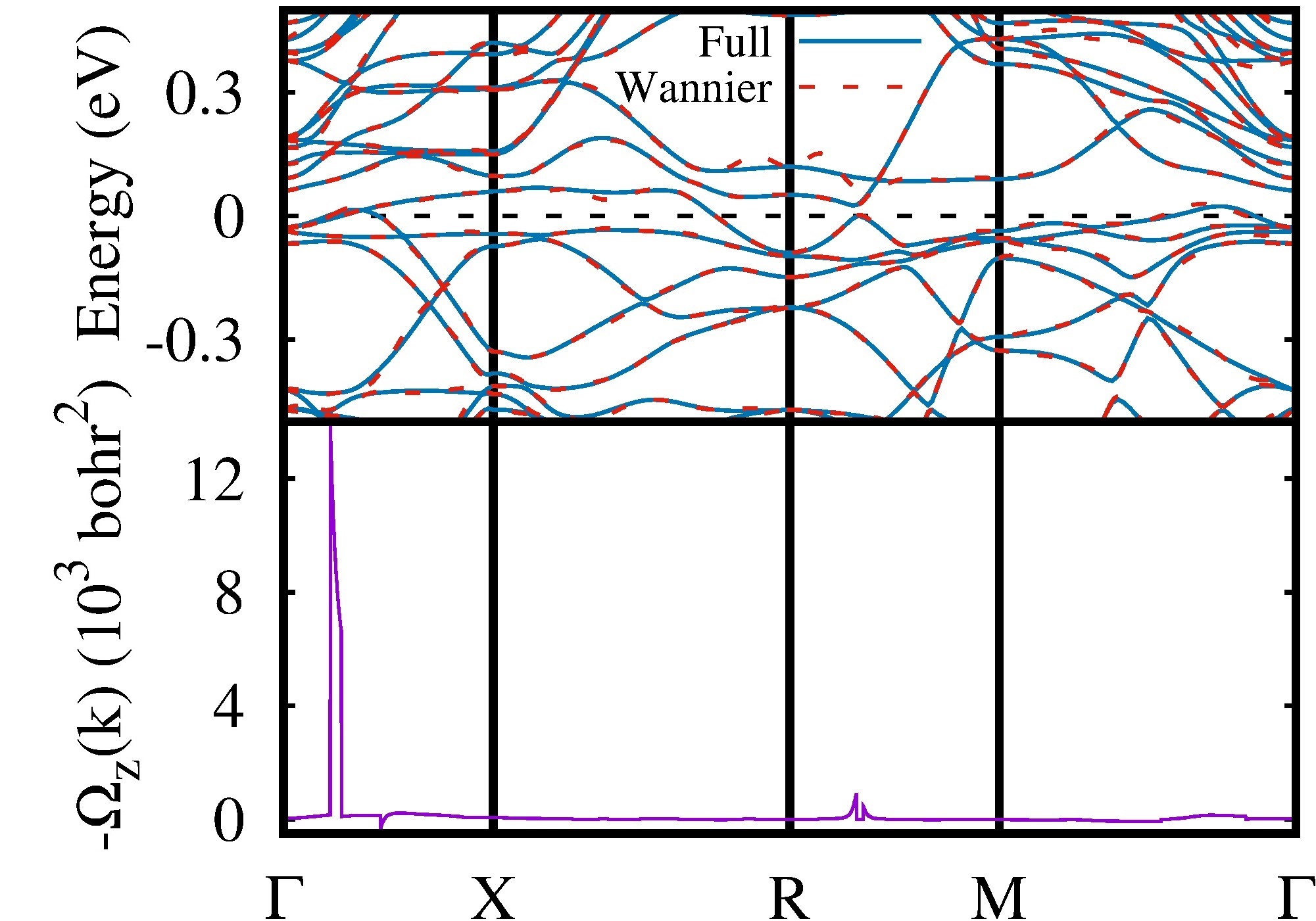}
\caption{Top: Full electronic band structure (blue continuous line) and Wannier interpolated band structure (red dashed line) of disordered Mn$_2$CoAl. The Fermi energy is set to 0 eV. Bottom: Distribution of the Berry curvature along high symmetry path in the Brillouin zone.}
\label{fig:disord}
\end{figure}

\begin{figure*}[t]
\centering
\includegraphics[width=1\linewidth]{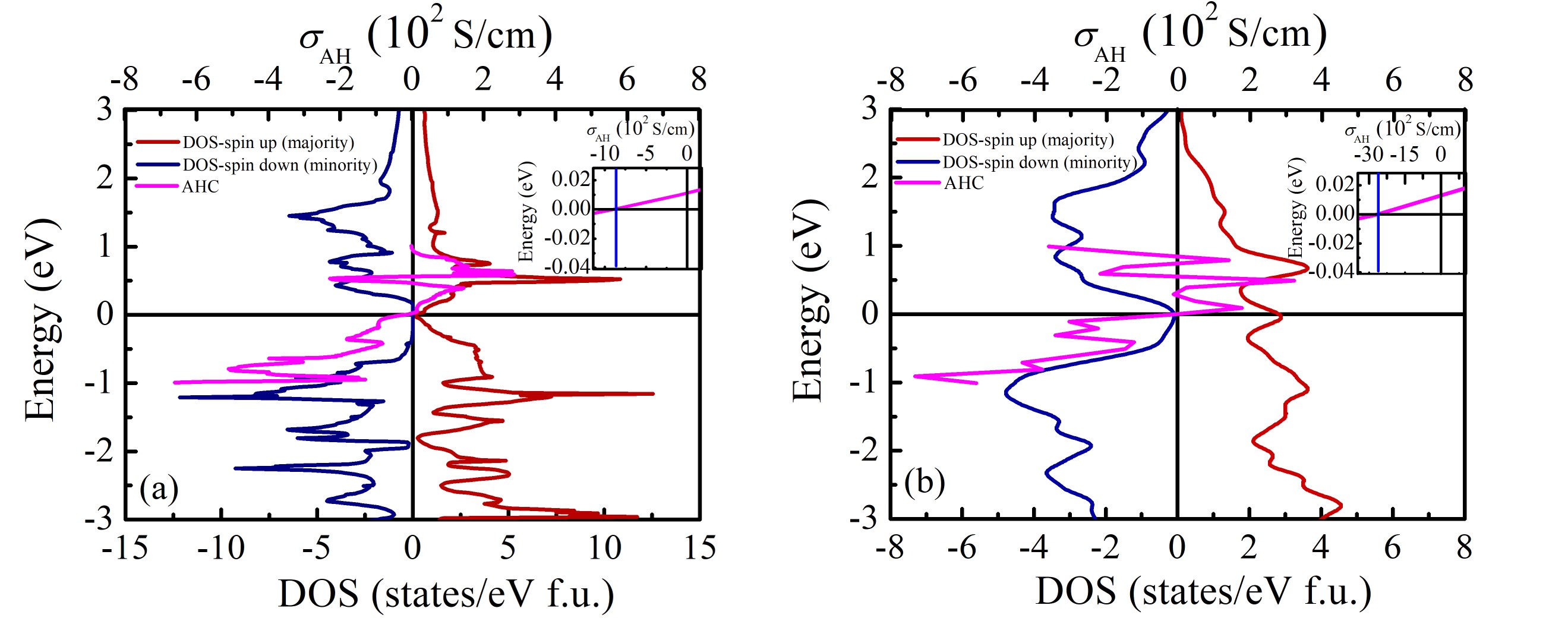}
\caption{(a) Density of states (DOS) and variation of intrinsic AHC of ordered Mn$_2$CoAl as function of the Fermi energy. The Fermi energy is set to 0 $eV$. (b) DOS and variation of intrinsic AHC for the disordered Mn$_2$CoAl as function of Fermi energy. The Fermi energy is set to 0 $eV$. Insets in (a) and (b) show an enlarged view of intrinsic AHC around Fermi energy.}
\label{fig:theory}
\end{figure*}

In literature, the value of AHC obtained from theory is very less as compared to the experiment \cite{ouardi2013}. Here, from the experiment, we found that the Mn$_2$CoAl has 25\% disorder between Mn\textsubscript{4c} and Al atoms. It is possible that this disorder may play an important role in the experimentally observed AHC, therefore we investigated the AHC in ordered and disordered Mn$_2$CoAl using theoretical calculations.

Before the discussion of AHC calculation, we first present our electronic structure results. The spin-polarized band structure of ordered Mn$_2$CoAl is shown in Fig.\,\ref{fig:mca-sp-wan_soc}. The spin-minority channel has a finite gap and the majority one has zero gap, which suggests SGS feature of the Mn$_2$CoAl compound. Spin gapless semiconductors offer intriguing transport properties because both the electrons and holes can be $100\%$ spin-polarized, so that spin can be controlled via only a small applied external energy. The band structure for up and down spins shown in Fig.\,\ref{fig:mca-sp-wan_soc} matches very well with the literature \cite{ouardi2013}. Our calculated magnetic moment for the stoichiometric Mn$_2$CoAl is found to be 2 \si{\mu}\textsubscript{B}/f.u. with anti-parallel coupling between Mn\textsubscript{4a} and Mn\textsubscript{4c} sites, which is also in agreement with the literature \cite{ouardi2013}. 

Now we will discuss the theoretical results on anomalous Hall transport obtained from Wannier90. We compared the full DFT band structure of ordered Mn$_2$CoAl with the Wannier interpolated one in Fig.\,\ref{fig:ord} (top) and obtained a very good interpolation. The distribution of calculated Berry curvature along the high symmetry path is shown in Fig.\,\ref{fig:ord} (bottom). 
An efficient first-principle approach has been used in which the maximally localized Wannier functions are first constructed from the Bloch states, on a relatively coarse $k$-grid. Then the quantities of interest e.g. Berry curvature is interpolated onto a dense $k$-mesh in calculating the intrinsic AHC as a Brillouin zone summation of the Berry curvature over all occupied states \cite{pizzi2020, kubler2012}- 

\begin{equation}
\sigma_{xy} (AHC) = \frac{e^2}{\hbar} \frac{1}{NV}\sum_{\textbf{k}\in(BZ)} (-1) \Omega_{xy}(\textbf{k}) f(\textbf{k}),
\end{equation}
\label{eq:ahc}

where the indices $x$ and $y$ are the Cartesian coordinates. $f(\textbf{k})$ is the Fermi distribution function, $\Omega_{xy}(\textbf{k})$ denotes the Berry curvature for the wave vector $\textbf{k}$, $N$ is the number of electrons in the crystal and $V$ is the cell volume.


Our theoretical calculation gives intrinsic AHC value \si{\sim}8.88 S/cm for the ordered Mn$_2$CoAl compound, which is slightly higher than the reported  theoretical intrinsic AHC value of 3 S/cm \cite{ouardi2013}. This difference could be due to two different approximations to compute AHC. Thus, the  theoretically calculated AHC for the ordered structure of the Mn$_2$CoAl compound is an order of magnitude smaller than the experimental value of intrinsic AHC \si{\sim}35 S/cm. Therefore, to understand this difference, we performed the theoretical calculations by incorporating the amount of anti-site disorder obtained from the SXRD experiment to compute the AHC. The magnetic moment obtained in this (disordered Mn$_2$CoAl) case also found to be \si{\sim}2 \si{\mu}\textsubscript{B}/f.u., which indicates that the anti-site disorder does not affect the magnetization and therefore, electronic structure is most likely deciding the observed AHC in disordered Mn$_2$CoAl. In Fig. \ref{fig:disord} (top), we have compared the full DFT band structure with the Wannier interpolated one for disordered structure. The distribution of calculated Berry curvature along the high symmetry path is shown in Fig. \ref{fig:disord} (bottom). Interestingly, the calculated intrinsic AHC in the disordered Mn$_2$CoAl increased to the 26.30 S/cm, which is close to our experimentally obtained intrinsic AHC. In the present case, the disordered sample still possesses certain space group symmetry, which suggests that the disorder in the atomic positions will also be repeated in the periodic units and the theoretical calculation also considers the same. It is true that the value in the real samples could be slightly different from the theoretical results. The possible source of significantly less discrepancy in the value of experimental and theoretical intrinsic AHC for the disordered Mn$_2$CoAl could be the involved approximations in all steps of the theoretical calculations (starting from obtaining ground state density to electronic and magnetic properties to Berry curvature computation to AHC).  
This suggests that the anti-site disorder modifies the Berry curvature and hence, enhances the intrinsic AHC in the Mn$_2$CoAl Heusler compound.


Recently,  a comparative study of AHC between SGS Mn$_2$CoGa and half-metallic Co$_2$VGa compounds has been performed, which reports that the Mn$_2$CoGa compound exhibits almost zero AHC, whereas half-metallic Co$_2$VGa (which has the similar magnetization and number of valence electrons as Mn$_2$CoGa) shows a larger AHC about 140 S/cm \cite{manna2018}. The origin of this behavior was explained based on the electronic structure of these two compounds. SGS compounds are characterized by a finite gap in the minority spin channel and a closed gap in the majority spin channel, which results in a few number of majority spin states around Fermi level. In the SGS compounds like Mn$_2$CoGa, the Berry curvature of majority spin states is reported to cancel out by the minority spin states resulting  into  a small/almost zero intrinsic AHC \cite{manna2018}. As compared to the SGS compounds, the half-metallic systems have a finite gap in minority spin channel and a large number of states in majority spin channel around the Fermi level \cite{rani2020}. In the half-metallic Co$_2$VGa compound, the large intrinsic AHC has been observed as compared to the Mn$_2$CoGa compound due to the uncompensated Berry curvature associated with a large number of majority spin states \cite{manna2018}. Our study follow this behavior as the ordered Mn$_2$CoAl exhibits the small intrinsic AHC due to its SGS nature (depicted in Fig,\,\ref{fig:theory}(a)), while with introducing the 25\% disorder the system seems to be half-metallic due to the formation of new majority spin states around the Fermi level, provides larger Berry curvature and leads to enhanced intrinsic AHC (depicted in Fig,\,\ref{fig:theory}(b)), which is in good agreement with the experiment. The inset of the Fig. \ref{fig:theory}(a) and (b) show an enlarged view of the intrinsic AHC value around the Fermi level for ordered and disordered Mn$_2$CoAl compounds, respectively. Thus, our combined  experimental theoretical results show that atomic disorder enhances the intrinsic AHC due to the modification in the Berry curvature linked with the electronic structure in the Mn$_2$CoAl compound. 

\section{Conclusion}
To conclude, we have presented here the evidence of anti-site disorder enhanced intrinsic AHC in the Mn$_2$CoAl Heusler compound by comprehensive analysis of the crystal structure and anomalous Hall effect using experimental and theoretical tools. The high resolution SXRD data reveals 25\% anti-site disorder between Mn\textsubscript{4c} and Al atoms within inverse Heusler structure. The temperature-dependent resistivity shows semiconducting behavior and follows Mooij's criteria for disordered metal. Scaling behavior suggests that the intrinsic mechanism dominates over the extrinsic mechanism in the AHE. The experimental intrinsic AHC is found to be larger than the theoretically reported value for the ordered Mn$_2$CoAl. The first-principle calculations conclude that the anti-site disorder enhances the Berry curvature induced intrinsic AHC, which is in well agreement with the experimentally found intrinsic AHC.

\section*{Acknowledgment}

We thankfully acknowledge UGC-DAE CSR, Indore for experimental support. SS thanks Science and Engineering Research Board of India for financial support through the award of Core Research Grant (grant no: CRG/2021/003256). NS thanks to DST INSPIRE scheme for financial support. Parts of this research were carried out at PETRA III of DESY, a member of the Helmholtz Association. Financial support from the Department of Science and Technology, Government of India within the framework of the India@DESY is thankfully acknowledged. We would like to show our gratitude towards the beamline scientist Dr. Martin Etter for his help in setting up the experiments. 

\par


\medskip

\begin{thebibliography}{}

\bibitem {book} C. M. Hurd, \textit{The Hall Effect in Metals and Alloys} (Plenum, New York, 1972).

\bibitem {nagaosa2010} N. Nagaosa, J. Sinova, S. Onoda, A. H. MacDonald, and N. P. Ong, Rev. Mod. Phys. {\bf82}, 1539 (2010).

\bibitem{smejkal2022} L. \v{S}mejkal, A. H. MacDonald, J. Sinova, S. Nakatsuji, and T. Jungwirth, Nat. Rev. Mater. (2022). 

\bibitem {friedmann2013} S. Friedemann, M. Brando, W. J. Duncan, A. Neubauer, C. Pfleiderer, and F M. Grosche, Phys. Rev. B {\bf87}, 024410 (2013).

\bibitem {golod2013} T. Golod, A. Rydh, P. Svedlindh, and V. M. Krasnov, Phys. Rev. B {\bf87}, 104407 (2013).

\bibitem {tian2009}  Y. Tian, L. Ye, and X. Jin, Phys. Rev. Lett. {\bf103}, 087206 (2009).

\bibitem {yu2010} R. Yu, W. Zhang, H.-J. Zhang, S.-C. Zhang, X. Dai, and Zhong Fang, Science {\bf329}, 61-64 (2010).

\bibitem{nayak2016} A. K. Nayak, J. E. Fischer, Y. Sun, B. Yan, J. Karel, A. C. Komarek, C. Shekhar, N. Kumar, W. Schnelle, J. K\"{u}bler, C. Felser, and S. S. P. Parkin, Sci. Adv. {\bf2}, e1501870 (2016).

\bibitem{chang2013} C.-Z. Chang, J. Zhang, X. Feng, Jie Shen, Z. Zhang, M. Guo, K. Li, Y. Ou, P. Wei, L.-L. Wang, Z.-Q. Ji, Y. Feng, S. Ji, X. Chen, J. Jia, X. Dai, Z. Fang, S.-C. Zhang, K. He, Y. Wang, L. Lu, X.-C. Ma, and Q.-K. Xue, Science {\bf340}, 167-170 (2013).

\bibitem{yang2020} S.-Y. Yang, Y. Wang, B. R. Ortiz, D. Liu, J. Gayles, E. Derunova, R. G.-Hernandez, L. $\Check{S}$mejkal, Y. Chen, S. S. P. Parkin, S. D. Wilson, E. S. Toberer, T. McQueen, and M. N. Ali, Sci. Adv. {\bf6}, eabb6003 (2020).

\bibitem{taniguchi2015} T. Taniguchi, J. Grollier, and M. D. Stiles, Phys. Rev. Appl. {\bf3}, 044001 (2015).

\bibitem{bauer2012} G. E. Bauer, E. Saitoh , and B. J. v. Wees, Nat. Mater. {\bf11(5)}, 391-9 (2012). 

\bibitem{kim2022} S. K. Kim, G. S. D. Beach, K. J. Lee, T. Ono, T. Rasing, and H. Yang, Nat. Mater. {\bf21}, 24–34 (2022). 

\bibitem{Jungwirth2016} T. Jungwirth, X. Marti, P. Wadley, and J. Wunderlich Nat. Nanotechnol. {\bf11}, 231–241 (2016).

\bibitem{smejkal2018}  L. \v{S}mejkal, Y. Mokrousov, B. Yan, and A. H. MacDonald Nat. Phys. {\bf14}, 242–251 (2018). 

\bibitem {bader2010} S.D. Bader, S.S.P. Parkin, Annu. Rev. Condens. Matter. Phys. {\bf1}, 71–88 (2010).
 
\bibitem {gambino1976} R. J. Gambino and T. Mcguire, IBM Tech. Disclosure Bull. {\bf18}, 4214 (1976).

\bibitem {gerber2007} A. Gerber, J. Magn. Magn. Mater. {\bf310}, 2749-2751 (2007).

\bibitem {moritz2008} J. Moritz, B. Rodmacq, S. Auffret, and B. Dieny,  J. Phys. D: Appl. Phys. {\bf41}, 135001 (2008).

\bibitem {karplus1954} R. Karplus and J. M. Luttinger, Phys. Rev. {\bf95}, 1145 (1954).

\bibitem {onoda2006} S. Onoda, N. Sugimoto, and N. Nagaosa, Phys. Rev. Lett. {\bf97}, 126602 (2006).

\bibitem {smit1958} J. Smit and J. Volger, Physica (Amsterdam) {\bf24}, 39 (1958).

\bibitem {berger1970} L. Berger, Phys. Rev. B {\bf2}, 4559-4566 (1970).

\bibitem {berry1984}  M. V. Berry, Proc. R. Soc. London {\bf392}, 45-57 (1984).

\bibitem {wang2007} X. Wang, D. Vanderbilt, J. R. Yates, and I. Souza, , Phys. Rev. B {\bf76}, 195109 (2007).

\bibitem {luttinger1958} J. M. Luttinger, Phys. Rev. {\bf112}, 739-751 (1958).


\bibitem{taguchi2001} Y. Taguchi, Y. Oohara, H. Yoshizawa, N. Nagaosa, and Y. Tokura, Science {\bf291}, 2573-2576 (2001).

\bibitem{jungwirth2002} T. Jungwirth, Q. Niu, and A. H. MacDonald, Phys. rev. lett., {\bf88}, 207208 (2002).

\bibitem {nakatsuji2015}  S. Nakatsuji, N. Kiyohara, and T. Higo, Nature {\bf527}, 212–215 (2015).


\bibitem{wang2016} Q. Wang, S. Sun, X. Zhang, F. Pang, and H. Lei, Phys. Rev. B {\bf94}, 075135 (2016).

\bibitem {wang2018} Q. Wang, Y. Xu, R. Lou, Nat. Commun. {\bf9}, 3681 (2018). 

\bibitem{ghimire2018}  N.J. Ghimire, A. S. Botana, J. S. Jiang, J. Zhang, Y. -S. Chen, and J. F. Mitchell, Nat. Commun. {\bf9}, 3280 (2018). 

\bibitem{yoo2021} M. W. Yoo, J. Tornos, A. Sander, L. F. Lin, N. Mohanta, A. Peralta, D. Sanchez-Manzano, F. Gallego, D. Haskel, J. W. Freel, D. J. Keavney, Y. Choi, J. Strempfer, X. Wang, M. Cabero, H. B. Vasili, M. Valvidares, G. Sanchez-Santolino, J. M. Gonzalez-Calbet, A. Rivera, C. Leon, S. Rosenkranz, M. Bibes, A. Barthelemy, A. Anane, E. Dagotto, S. Okamoto, S. G. E. Te Velthuis, J. Santamaria, J. E. Villegas, Nat. Commun. {\bf12}, 3283 (2021). 

\bibitem{ernst2019} B. Ernst, R. Sahoo, Y. Sun, J. Nayak, L. Müchler, A. K. Nayak, N. Kumar, J. Gayles, A. Markou, G. H. Fecher, and C. Felser, Phys. Rev. B {\bf100}, 054445 (2019).

\bibitem{manna2018} K. Manna, L. Muechler, T. H. Kao, R. Stinshoff, Y. Zhang, J. Gooth, N. Kumar, G. Kreiner, K. Koepernik, and R. Car, J. K\"{u}bler, G. H. Fecher, C. Shekhar, Y. Sun, and C. Felser, Phys. Rev. X {\bf8}, 041045 (2018).


\bibitem {Li2020} P. Li, J. Koo, W. Ning, J. Li, L. Miao, L. Min, Y. Zhu, Y. Wang, N. Alem, C. X. Liu, Z. Mao, and B. Yan, Nat. Commun. {\bf11}, 3476 (2020).

\bibitem{gaurav2021} G. K. Shukla, J. Sau, N. Shahi , A. K. Singh, M. Kumar, and S. Singh, Phys. Rev. B {\bf104}, 195108 (2021). 

 \bibitem {Noky2020} J. Noky, Y. Zhang, J. Gooth, C. Felser, and Y. Sun, Npj Comput. Mater. {\bf6}, 77 (2020).

 \bibitem {shen2020}  J. Shen, Q. Yao, Q. Zeng, H.  Sun, X.  Xi, G. Wu, W. Wang, B. Shen, Q. Liu, and E. Liu, Phys. Rev. Lett. {\bf125}, 086602 (2020).
 
\bibitem{vidal2011} E. V. Vidal, H. Schneider, and G. Jakob, Phy. Rev. B {\bf83}, 174410 (2011).


\bibitem{hazra2018} B. K. Hazra, M. M. Raja, R. Rawat, A. Lakhani, S. Kaul, and S. Srinath, J. Magn. Magn.
Mater. {\bf448}, 371 (2018).

\bibitem {wang2019} Z. Wang, Q.  Liu, J. W. Luo, and A. Zunger, Mater. Horiz. {\bf6}, 2124 (2019).
   
 \bibitem {kastbjerg2013} S. Kastbjerg, N. Bindzus, M. Søndergaard, S. Johnsen, N. Lock, M. Christensen, M. Takata, M. A. Spackman, and B. B. Iversen, Adv.  Funct.  Mater. {\bf23}, 5477–5483 (2013).

  \bibitem{mende2021} F. Mende, J. Noky, S. N. Guin, G. H. Fecher, K. Manna,P. Adler, W. Schnelle, Y. Sun, C. Fu, and C. Felser, Adv. Sci. {\bf8}, 2100782 (2021).

 \bibitem{gauravc2021} G. K. Shukla, A. K. Jena, N. Shahi, K. K. Dubey, I. Rajput, S. Baral, K. Yadav, K. Mukherjee, A. Lakhani, K. Carva, S.-C. Lee, S. Bhattacharjee, S. Singh, Phys. Rev. B {\bf105}, 035124 (2022).
 
 \bibitem{roy2020} S. Roy , R. Singha, A. Ghosh, A. Pariari, and P. Mandal, Phys. Rev. B. {\bf102}, 085147 (2020).


 
 
 

 
 
  \bibitem{ouardi2013} S. Ouardi, G. Fecher, C. Felser, and J. K\"{u}bler, Phys. Rev. Lett. {\bf110}, 100401 (2013).
  
  
  \bibitem{wang2008} X. L. Wang, Phys. Rev. Lett., {\bf100}, 156404 (2008).

  \bibitem{arima2018} K. Arima, F. Kuroda, S. Yamada, T. Fukushima, T. Oguchi, and K. Hamaya, Phys. Rev. B {\bf97}, 054427 (2018).
  
  \bibitem{buckley2019} R. G. Buckley, T. Butler, C. Pot, N. M. Strickland and S. Granville, Mater. Res. Express {\bf6}, 106113 (2019).
  
  \bibitem{xu2019} X. D. Xu, Z. X. Chen, Y. Sakuraba, A. Perumal, K. Ma-suda, L. S. R. Kumara, H. Tajiri, T. Nakatani, J. Wang, W. Zhou, Y. Miura, T. Ohkubo, and K. Hono, Acta Mater. {\bf176}, 33-42 (2019).
  
   \bibitem{chen2018} P. Chen, C. Gao, G. Chen, K. Mi, M. Liu, P. Zhang, and D. Xue, Appl. Phys. Lett. {\bf113}, 122402 (2018).
   
 \bibitem{feng2015} Y. Feng, C.-l. Tian, H.-k. Yuan, A.-l. Kuang and H. Chen, J. Phys. D: Appl. Phys. {\bf48}, 445003 (2015).
 
 \bibitem{sun2016} N. Y. Sun, Y. Q. Zhang, H. R. Fu, W. R. Che, C. Y. You, and R. Shan, AIP Adv. {\bf6}, 015006 (2016).
 
 \bibitem{Xin2017} Y. Xin, H. Hao, Y.Ma, H. Luo, F. Meng, H. Liu, E. Liu, and G. Wu, Intermetallics {\bf80}, 10-15 (2017).
  
 \bibitem{zhang2013} Y. J. Zhang, G. J. Li, E. K. Liu, J. L. Chen, W. H. Wang, and G. H. Wu, J. Appl. Phys. {\bf113}, 123901 (2013).
 
  \bibitem{xu2014} G. Z. Xu, Y. Du, X. M. Zhang, H. G. Zhang, E. K. Liu, W. H. Wang, and G. H. Wu, Appl. Phys. Lett. {\bf104}, 242408 (2014).
  
   \bibitem{jamer2013} M. E. Jamer, B. A. Assaf, T. Devakul, and D. Heiman, App. Phys. Lett. {\bf103}, 142403 (2013).
  
  \bibitem{ueda2017} K. Ueda, S. Hirose, and H. Asano, App. Phys. Lett. {\bf110}, 202405 (2017).
  
   \bibitem{kudrnovsky2013} J. Kudrnovsk\'{y}, V. Drchal, and I. Turek, Phys. Rev. B {\bf88}, 014422 (2013).
   
\bibitem{holt1978} B. Holt, J. Diaz, J. Huber, and C. A. Luengo, Rev. Bras. Ensino Fis. {\bf8}, 155-163 (1978).

\bibitem{sanjay2017} S. Singh, B. Dutta, S. W. D’Souza, M. G. Zavareh, P. Devi, A. S. Gibbs, T. Hickel, S. Chadov, C. Felser, and D. Pandey, Nat Commun {\bf8}, 1006 (2017).

\bibitem{sanjay2012} S. Singh, R. Rawat, S. E. Muthu, S. W. D’Souza, E. Suard, A. Senyshyn, S. Banik, P. Rajput, S. Bhardwaj, A. M. Awasthi, R. Ranjan, S. Arumugam, D. L. Schlagel, T. A. Lograsso, A. Chakrabarti, and S. R. Barman, Phys. Rev. Lett. {\bf109}, 246601 (2012).

   
  \bibitem{giannozzi2009} P. Giannozzi, S. Baroni, N. Bonini, M. Calandra, R. Car, C. Cavazzoni, D. Ceresoli, G. L. Chiarotti, M. Cococcioni, I. Dabo, A. D. Corso, S. de Gironcoli, S. Fabris, G.  Fratesi, R. Gebauer, U. Gerstmann, C. Gougoussis,A. Kokalj, M. Lazzeri, L. Martin-Samos, N. Marzari, F. Mauri, R. Mazzarello, S. Paolini, A. Pasquarello, L. Paulatto, C. Sbraccia, S. Scandolo, G. Sclauzero, A. P.Seitsonen, A. Smogunov, P. Umari, and R. M. Wentzcovitch, J. Phys. Cond. Matt. {\bf21}, 395502 (2009).

\bibitem{perdew1996} J. P. Perdew, K. Burke, and M. Ernzerhof, Phys. Rev. Lett. {\bf77}, 3865 (1996).

\bibitem{hamann2013} D. R. Hamann, Phys. Rev. B {\bf88}, 085117 (2013).

\bibitem{marzari2001} N. Marzari and D. Vanderbilt, Phys. Rev. B {\bf56}, 12847(1997).

\bibitem{souza2001} I. Souza, N. Marzari, and D. Vanderbilt, Phys. Rev. B {\bf65}, 035109 (2001).

\bibitem{pizzi2020} G. Pizzi, V. Vitale, R. Arita, S. Bl\"{u}gel, F. Freimuth, G. G\'{e}ranton, M. Gibertini, D. Gresch, C. Johnson, T. Koretsune, J. I.-Azpiroz, H. Lee, J.-M. Lihm, D. Marchand, A. Marrazzo, Y. Mokrousov, J. I. Mustafa, Y. Nohara, Y. Nomura, L. Paulatto, S. Poncé, T. Ponweiser, J. Qiao, F. Th\"{o}le, S. S. Tsirkin, M. Wierzbowska, N. Marzari, D. Vanderbilt, I. Souza, A. A. Mostofi, and J. R. Yates, J. Phys. Cond. Matt. {\bf32}, 165902 (2020).

\bibitem{graf2011} T. Graf, C. Felser, and S. S. P. Parkin, Prog. Solid. State Ch. {\bf39}, 1-50 (2011).

\bibitem{full} J. Rodr\'{i}guez-Carvajal, FULLPROF, a Rietveld and pattern matching and analysis programs version 2016, Laboratoire Leon Brillouin, CEA-CNRS, France,
\bibinfo{link}{\textcolor{blue}{http://www.ill.eu/sites/ fullprof/}}

\bibitem{nayak2015} A. Nayak, M. Nicklas, S. Chadov, P. Khuntia, C. Shekhar, A. Kalache, M. Baenitz, Y. Skourski, V. K. Guduru, A. Puri, U. Zeitler, J. M. D. Coey, and C. Felser, Nat. Mat. {\bf14}, 679–684 (2015).

\bibitem{nayak2013} A. K. Nayak,  M. Nicklas, S. Chadov, C. Shekhar, Y. Skourski, J. Winterlik, and C. Felser, Phys. Rev. Lett. {\bf110}, 127204 (2013).

\bibitem{dliu2008} G. D. Liu, X. F. Dai, H. Y. Liu, J. L. Chen, Y. X. Li, G. Xiao, and G. H. Wu, Phys. Rev. B {\bf77}, 014424 (2008).



\bibitem{momma2011} K. Momma and F. Izumi, \enquote{VESTA 3 for three-dimensional visualization of crystal, volumetric and morphology data}, J. Appl. Crystallogr. {\bf44}, 1272-1276 (2011).


\bibitem{bezmat2014} L. N. Bezmaternykh, E. M. Kolesnikova, E. V. Eremin, S. N. Sofronova, N.V. Volkov, and M.S. Molokeev, J. Magn. Magn {\bf364}, 55-59 (2014).


 \bibitem{galanakis2014} I. Galanakis, K. \"{O}zdoan, E. \,{S}a\,{s}ioglu, and S. Bl\"{u}gel, J. Appl. Phys. {\bf115}, 093908 (2014).

\bibitem{dugdale1995} S. Dugdale, The Electrical Properties of Disordered Metals (Cambridge University Press, Cambridge)  (1995).

\bibitem{hikami1980} S. Hikami, A. I. Larkin, and Y. Nagaoko, Prog. Theor. Phys. {\bf63}, 707 (1980).

\bibitem{xu2014} G. Xu, W. Wang, X. Zhang, Y. Du, E. Liu, S. Wang, G. Wu, Z. Liu, and X. X. Zhang, Sci. Rep. {\bf4}, 5709 (2014).

\bibitem {Singh2022} M. Singh, L. Ghosh, V. K. Gangwar, Y. Kumar, D. Pal, P.  Shahi, S. Kumar, S. Mukherjee, K. Shimada, and S. Chatterjee, Appl. Phys. Lett. {\bf121}, 032403 (2022).

\bibitem{chamorro2019} J. R. Chamorro, A. Topp, Y. Fang, M. J. Winiarski, C. R. Ast, M. Krivenkov, A. Varykhalov, B. J. Ramshaw, L. M. Schoop, and T. M. McQueen, APL Mater. {\bf7}, 121108 (2019).

\bibitem{sasmal2020} S. Sasmal, R. Mondal, R. Kulkarni, A. Thamizhavel, and B. Singh, J. Phys. Condens. Matter {\bf32}, 335701 (2020).

\bibitem{laha2019} A. Laha, S. Malick, R. Singha, P. Mandal, P. Rambabu, V. Kanchana, and Z. Hossain, Phys. Rev. B {\bf99}, 241102 (2019).

\bibitem{chen2020} J. Chen, H. Li, B. Ding, Z. Hou, E. Liu, X. Xi, G. Wu, and W. Wang, Appl. Phys. Lett. {\bf116}, 101902 (2020).

\bibitem{dulal2019} R. P. Dulal, B. R.Dahal, A. Forbes, N. Bhattarai, I. L. Pegg, and J. Phili, Sci Rep {\bf9}, 3342 (2019).

\bibitem{kawabata1980} A. Kawabata, J. Phys. Soc. Jpn. {\bf49}, 628-637 (1980)

\bibitem{baxter1989} D. V. Baxter, R. Richter, M. L. Trudeau, R. W. Cochrane, and J. O. Strom-Olsen, J. Phys. France {\bf50}, 1673-1688 (1989). 

\bibitem{chen2019} , W. Chen, H.-Z. Lu, and O. Zilberberg, Phys. Rev. Lett. {\bf122}, 196603 (2019).
  
\bibitem{malick2022} S. Malick, A. Ghosh, C. K. Barman, A. Alam, Z. Hossain, P. Mandal, and J. Nayak, Phys. Rev. B {\bf105}, 165105 (2022).


\bibitem{afzal2020} H. Afzal, S. Bera, A.K. Mishra, M. Krishnan, M. M. Patidar, R. Venkatesh, and V. Ganesan, J. Supercond. Nov. Magn., 1-8 (2020).

\bibitem{mooij1973} J. H. Mooij, Phys. Stat. Sol. (a) {\bf17}, 521 (1973).


 
 \bibitem{kaveh1982} M. Kaveh, and N. F. Mott, J. Phys. C: Solid State Phys. {\bf15}, L707 (1982).
 
 \bibitem{gofryk2005} K. Gofryk, D. Kaczorowski, and T. Plackowski, Phys. Rev. B {\bf72}, 094409 (2005).
 
  \bibitem{kida2011} T. Kida, L. A. Fenner, A. A. Dee, I. Terasaki, M. Hagi-wara1, and A. S. Wills, J. Phys.:  Condens. Matter {\bf23}, 112205 (2011).
  
  \bibitem{he2012} P. He, L. Ma, Z. Shi, G. Y. Guo,  and J.-G. Zheng, and Y. Xin, and S. M. Zhou, Phys. Rev. Lett. {\bf109}, 066402 (2012).

  \bibitem{kotzler2005} J. K\"otzler, and W. Gil, Phys. Rev. B, {\bf72},060412 (2005).
  
  \bibitem{nozieres1973} P. Nozieres and C. Lewiner, Journal de Physique {\bf34}, 901 (1973).

\bibitem{onoda2006} S. Onoda, N. Sugimoto, and N. Nagaosa, Phys. Rev. Lett. {\bf97}, 126602 (2006).

\bibitem{kim2018} K. Kim, J. Seo, E. Lee, K.-T. Ko, B. S. Kim, B. G. Jang, J. M. Ok, J. Lee, Y. J. Jo, W. Kang, J. H. Shim, C. Kim, H. W. Yeom, B. I. Min, B.-J. Yang, and J. S. Kim, Nat. Mater. {\bf17}, 794–799 (2018).
 
 
 \bibitem{zeng2006} C. Zeng, Y. Yao, Q. Niu, and H. H. Weitering, Phys. Rev. Lett. {\bf96}, 037204 (2006).
 
\bibitem{eliu2018} E. Liu, Y. Sun, N. Kumar, L. Muechler, A. Sun, L. Jiao,S. Y. Yang, D. Liu, A. Liang, Q. Xu, J. Kroder, V. S\"{u}ß, H. Borrmann, C. Shekhar, Z. Wang, C. Xi, W. Wang, W.Schnelle, S. Wirth, Y. Chen, S. T. B. Goennenwein, and C. Felser, Nat. Phys. {\bf14}, 1125–1131 (2018). 



 \bibitem{kubler2012} J. K\"{u}bler and C. Felser, Phys. Rev. B {\bf85}, 012405 (2012).
 
\bibitem{rani2020} D. Rani, L. Bainsla, A. Alam, K. G. Suresh, Journal of Applied Physics {\bf128}, 220902 (2020).
 
 
  






 
  \end{thebibliography}
\end{document}